\documentstyle[12pt]{article}
\begin{document}

\title{The interaction of dyons in the mean field approximation }
\author{B.V.Martemyanov \\
\it {Institute of Theoretical} $\&$ \it {Experimental Physics}\\
\it {117259, B.Cheremushkinskaya 25, Moscow, Russia}}

\maketitle
\begin{abstract}
The interaction of dyons in the mean field approximation is considered.
The result of interaction is the mass term for dyonic field in the effective
Lagrangian. Due to the mass term the profile function of dyon falls off
exponentially at large distances.
\end{abstract}
\section{Introduction}
There is a hope that dyons, euclidean solutions of gauge theory,
can be the fluctuations of vacuum fields that are responsible for
confinement [1-4]. If we consider the gas of dyons (the simple
superposition of dyonic solutions), neglect the interaction of dyons
and omit their possible interference in the contribution to the
Wilson loop we will get the phenomenon of "superconfinement" [2-4].
The "superconfinement" means that the Wilson loop average $W$ follows
not the area low (confinement)
$$
W \sim exp(-Cr^2)
$$
where $C$ is some constant, $r$ is the radius of the loop
(the circle, for example), but falls off like $exp(-C'r^3)$
($C'$ is the constant). It was put forward the idea [2-4]
that due to the interaction of dyons something like "Debye
screening" should appear in the dyonic gas transforming the
"superconfinement" to confinement. In this paper we will
consider the interaction of dyons in the mean field approximation
and argue that effectively at large distances the dyonic field is
exponentially damped. Such a damping is known for a long time in the case
of instanton gas model of vacuum [5], where the damping was obtained
from the Feynmann variational principle. We will consider the connection
of the mean field approximation to the Feynmann variational principle.
Now we are not in a position to say whether this damping is the
desired "Debye screening" or not. In this sense the problem needs
further investigation.
\section{Mean field approximation}
Let us consider the gas of dyons: $N$ dyons in the volume $V$.
Dyons are described by their degrees of freedom: positions,
color orientations (we consider $SU(2)$ gauge theory), "velocities".
We consider one isolated dyon in the field of other dyons and average
the total action over their degrees of freedom. In this way we can
obtain the effective action for the dyon. Taking the minimum of the
effective action we can find the modified dyonic solution. And at
last we can use this modified solution to calculate some parameters
of the effective action.\\
The described procedure is usually called the mean field
approximation. We will show further the connection of this procedure
with some form of the Feynmann variational principle. And now
let us describe the details.\\
The total gauge field $A_\mu (x) = a_\mu (x) + B_\mu (x)$,
where $a_\mu (x)$ is the gauge field of the isolated dyon and
$B_\mu (x)$ is the field created by other dyons. The total action
is equal to
\begin{equation}
S = \int d^4 x \frac{1}{4} (f^a_{\mu\nu} + F^a_{\mu\nu} +
f^{int,a}_{\mu\nu})^2,
\end{equation}
where $f^{int,a}_{\mu\nu} = g\epsilon^{abc}(a_\mu^b B_\nu^c +
B_\mu^b a_\nu^c)$, $f_{\mu\nu}$ and $F_{\mu\nu}$ are field
tensors for potentials $a_\mu$ and $B_\mu$ respectively.
The averaging of $S$ over the field $B_\mu$ is rather simple
and as a result we get the effective action ($S_{eff}$) for the
isolated dyon
\begin{equation}
S_{eff} = \int d^4 x (\frac{1}{4} f_{\mu\nu}^2 + \frac{m^2}{2}
a_\mu^2) + S_B,
\end{equation}
where $m^2 = \frac{g^2}{2}<B_\mu^2>, S_B = \frac{1}{4}
<F_{\mu\nu}^2>$.  The constant $S_B$ is inessential for the
calculation of the modified one-dyon solution.\\ The effective action
is no more guage invariant and this is the reflection of the fact
that the interaction of dyons depend in the model (simple
superposition of individual dyons) on the gauge used for individual
dyonic field [2]. The problem here is that the superposition of
dyon's potentials is not only the solution of Yang-Mills equations
but also gives the resulting field tensor depending on the gauge
we are using for the dyon solution (the sum of potentials in
one gauge is not connected to that in another gauge by any gauge
transformation). The gauge should be taken in such a form as to get
a minimal interaction of dyons (at large distances, at least). The
interaction of dyons at large distances depends crucially on the
asymptotic behaviour of dyonic potential and the latter depends on
the class of the gauge, the different classes being connected by
singular gauge transformations.
In ref.[2] it was pointed out that the interaction of two dyons
vanishing at infinity
can
be obtained in the so
called 't Hooft gauge. We will use this gauge here also.
We use the 't Hooft gauge ($\partial_\mu a_\mu = 0$)
for the individual dyon solution in the superposition ansatz
also for the following reasons.  First, the topological charges
of the dyons are summed in the considered (singular) gauge.
Second, the interaction of dyons is minimal in this gauge
(at least at infinitisimal gauge transformation). And third,
because of the screened (due to the mass term) dyon solution
satisfies the condition $\partial_\mu a_\mu = 0$
automatically and transforms to the unscreened solution in the
$m^2 \rightarrow 0$ limit, the unscreened solution should
satisfy the condition $\partial_\mu a_\mu = 0$ also.

In the 't
Hooft gauge the unscreened dyon solution is equal to $$
ga_i^a=-\epsilon _{iab}n_bf(r,t) + \delta _i^ag(r,t) $$ $$ ga_0^a=-
n^af(r,t) $$ \begin{equation} f(r,t)=\frac 1r+\gamma
(\frac{\sinh\gamma r}{\cosh\gamma r-\cos\gamma t}-\frac{\cosh\gamma
r}{\sinh\gamma r}) \end{equation} $$ g(r,t)=\gamma \frac{\sin\gamma
t}{\cosh\gamma r-\cos\gamma t}. $$ Here $\gamma$ is the inverse size
of the dyon. The mass term in the effective action (2) modifies the
solution (3). If $m << \gamma $ (the limit of pointlike dyons)
qualitatively we can say that the internal part of the dyon is mainly
unchanged (the function $g$ and the last two terms of function $f$ in
(3) are unchanged) and the first term of function $f$ (Coulomb-like
tail) $\frac{1}{r}$ transforms to $\frac{exp(-mr)}{r}$ at $r >>
\frac{1}{m}$. So, we have obtained the exponential damping of the
dyonic solution at large distances. Obviously, the problem needs
further numerical consideration at this point. The parameter $m$ of
the effective action (2) can be now calculated using the modified
one-dyon solution.\\ According to eqs.(2),(3) $$m^2 =
(N-1)\frac{1}{2} g^2<a_\mu^2>=$$ \begin{equation}
=\frac{(N-1)}{V}4\pi\frac{3}{2} \int r^2dr (f^2 + g^2).
\end{equation}
In the limit of pointlike dyon we have approximately
\begin{equation}m^2 \approx n 4\pi \frac{3}{2} \frac{1}{2m}, (m <<
\gamma),\end{equation}
where $n =\frac{N}{V}$ is 3-d density of dyons.\\
 For the unscreened solution the integral in
formular (4) is divergent. Eq.(5) has the selfconsistent solution
\begin{equation}
m = ^3\sqrt{3\pi n}
\end{equation}

\section{The connection to the Feynmann variational principle.}
The modification of the dyon solution can be also found by Feynmann
variational principle. Let us assume that we are calculating the
partition function $Z$ for $N$ dyons:
\begin{equation}
Z = \int exp(-S),
\end{equation}
where the integral goes over the dyons degrees of freedom.
According to Feynmann variational principle
\begin{equation}
Z \ge Z_1 exp(-<S-S_1>),
\end{equation}
where $S_1$ and $Z_1$ are the simplified action and the
partition function. If we choose the action $S_1$ as a sum of
$N$ independent (no interaction) terms: $S_1 = Ns_1$ then
$$
Z_1 = \int exp(-Ns_1) = Aexp(-Ns_1)
$$
\begin{equation}
<S_1> = Ns_1,
\end{equation}
where $A$ is the volume of the space of dyons degrees of freedom.
So, we get
\begin{equation}
Z \ge A exp(-<S>).
\end{equation}
For $N$ dyons $<S>$ is equal to
\begin{equation}
<S> = N\int d^4 x \frac{1}{4} f_{\mu\nu}^2 + \frac{1}{4}
\frac{N(N-1)}{2} g^2<a_\mu^2> \int d^4 x a_\mu^2
\end{equation}
The variation of $<S>$ over $a_\mu$ (we are looking for the
maximum of $Z$ and minimum of $<S>$) gives the same equation as the
variation of $S_{eff}$ over $a_\mu$. The mass parameter $m^2$ is
equal to
$$m^2 = (N-1)\frac{1}{2} g^2<a_\mu^2>$$
in agreement with the result (4) obtained in the mean field
approximation. We have applied the mean field approximation to
the case of instantons (with fixed size $\rho$ for example).
The mass term for the instanton field calculated in such a way
coinsides with that of ref.[5] where the same problem was considered
in the context of Feynmann variational principle.
\section{Conclusion}
We have considered the problem of the dyons interaction in the
mean field approximation. In this approximation the interaction
is effectively taken into account in the form of mass term
for the dyonic field. The mass parameter ($m$) is selfconsistently
determined by the density of dyons (see eq.(5)). If other
fluctuations (not of dyonic type) are present in the vacuum
they also contribute to $m^2$. So, $m$ is larger than $^3\sqrt{3\pi
n}$. The effect of the considered mass term is the exponential
damping of the dyonic field at large distances.

This work is partly supported by the  RFFI grants 96-02-00088G
and 97-02-17491.
The author would like to thank Yu.A.Simonov for stimulating
discussions.

{\bf References}\\
1. Yu.A.Simonov, {\it Sov.J.Yad.Fiz.} {\bf 43} (1985) 557.\\
2. Martemyanov B.V., Molodtsov S.V., Simonov Yu.A. and
Veselov A.I., {\it Pis'ma Zh.Eksp.Teor.Fiz.} {\bf 62} (1995) 695\\
3. Martemyanov B.V., Molodtsov S.V.,{\it Sov.J.Yad.Fiz.} {\bf 59}
(1996) 766.\\
4. Martemyanov B.V., Molodtsov S.V., Simonov Yu.A. and
Veselov A.I.,{\it Sov.J.Yad.Fiz.} (1997) (to be published).\\
5. D.I.Diakonov and
V.Yu.Petrov, {\it Nucl.Phys.} {\bf B245} (1984) 259.

\end{document}